\documentclass[
 reprint,
 amsmath,amssymb,
 aps,prl
]{revtex4-2}

\usepackage{graphicx}
\usepackage{dcolumn}
\usepackage{bm}
\usepackage{xcolor}
\usepackage{gensymb}
\usepackage[colorlinks=true,bookmarks=false,citecolor=blue,urlcolor=blue]{hyperref}

\begin{document}

\preprint{APS/123-QED}

\title{Bound States in the Continuum in Multipolar Lattices}% Force line breaks with \\

\author{Sergei Gladyshev}%
\author{Artem Shalev}
\author{Kristina Frizyuk}%
\author{Konstantin Ladutenko}%
\author{Andrey Bogdanov}%
 \email{a.bogdanov@metalab.itmo.ru}
\affiliation{%
 School of Physics and Engineering, ITMO University, 191002, St. Petersburg, Russia}
\date{\today}% It is always \today, today,

\begin{abstract}

We develop a theory of bound states in the continuum (BICs) in multipolar lattices -- periodic arrays of resonant multipoles. We predict that BICs are completely robust to changes in lattice parameters remaining pinned to specific directions in the $k$-space. The lack of radiation for BICs in such structures is protected by the symmetry of multipoles forming the lattice.  We also show that some multipolar lattices can host BICs forming a continuous line in the $k$-space and such BICs carry zero topological charge. The developed approach sets a direct fundamental relation between the topological charge of BIC and the asymptotic behavior of the Q-factor in its vicinity. We believe that our theory is a significant step towards gaining deeper insight into the physics of BICs and the engineering of high-Q states in all-dielectric metasurfaces.         
\end{abstract}

\maketitle

{\it Bound states in the continuum} (BICs) are non-radiating solutions of the wave equation with the spectrum embedded in the continuum of the propagating modes in the surrounding space~\cite{hsu2016bound}. A general wave phenomenon, BICs were first predicted in quantum mechanics ~\cite{von1993merkwurdige} but today have found a variety of applications in photonics and acoustics~\cite{azzam2021photonic,sadreev2021interference,joseph2021bound,koshelev2019meta,koshelev2019nonradiating}. Their strong spatial localization and high quality (Q) factor provide giant amplification of the external electric field~\cite{yoon2015critical} and drastically enhance the light-matter interaction~\cite{qin2021strong,koshelev2018strong,kravtsov2020nonlinear,dyakov2021photonic}, nonlinear optical effects~\cite{koshelev2020subwavelength,sinev2021observation,koshelev2019nonlinear,zograf2022high}, and the performance of lasers~\cite{mylnikov2020lasing,wu2020room,wang2021highly,kodigala2017lasing,hwang2021ultralow,yang2021low} and optical biosensors~\cite{romano2018label,wang2021all,tittl2018imaging,maksimov2020refractive}.

One of the most used platforms supporting BICs is the periodic photonic structures including gratings~\cite{bulgakov2018propagating}, chains~\cite{sidorenko2021observation}, corrugated waveguides~\cite{hemmati2019resonant}, metasurfaces~\cite{cong2019symmetry} and photonic crystal slabs~\cite{zhen2014topological}.
The periodicity makes radiation possible only through open diffraction channels of which there is a finite number as opposed to a single scatterer where there is an infinite number. To form the BIC one needs to nullify the coupling constants to all of these channels which can be achieved by either  exploiting the symmetry of the structures or by precise tuning of the system's parameters, referred to as {\it symmetry-protected} and {\it parametric}, respectively~\cite{hsu2013observation}. In structures with subwavelength period, there is only one open diffraction channel that makes the engineering and observation of BICs substantially easier.  The symmetry-protected BICs  appear in the center of the Brillouin zone ($\Gamma$-point), while the parametric BICs appear out of the $\Gamma$-point, in the general case.

Dielectric and plasmonic metasurfaces have a strong optical response, usually associated with Mie or plasmonic resonances of meta-atoms~\cite{cherqui2019plasmonic,tonkaev2020high}. Each meta-atom possesses a certain multipolar content that depends on the symmetry of the unit cell and design of the meta-atom~\cite{chen2019singularities,sadrieva2019multipolar,gladyshev2020symmetry}. However, in the vicinity of the resonances, there usually is only one dominant multipole. Thus, the metasurfaces can be effectively considered as {\it multipolar lattices}, consisting of particular point multipoles~\cite{kostyukov2021multipolar,abujetas2020coupled,abujetas2022tailoring}. Multipolar lattices extend the concept of single electromagnetic multipoles which proved to be a powerful tool for describing optical properties of single resonant scatterers and nanoantennas~\cite{alaee2019exact,santiago2019decomposition,krasnok2012all}.   

% \red{thats a copy of the abstract? maybe}
% In this Letter we develop a theory of {\it bound states in the continuum in multipolar lattices}. We show that off-$\Gamma$ BICs can be {\it angular pinned} and are completely robust to the change of lattice period and resonant frequency of the multipoles. These BICs can form a continuous or discontinuous line in $k$-space but not a point as usual for multipole with azimutal number $m=0$ and order $l>2$.
% The topological properties of BIC in a multipolar lattice are studied. We demonstrate that off-$\Gamma$ BICs with a continuous or discontinuous line in $k$-space have the zero topological charge. The theory give a direct fundamental relation between the topological charge of $\Gamma$ BIC and the asymptotic behavior of Q-factor of the radiative modes in its vicinity. And we have demonstrate the formula of Q-factor depending on k-vector for multipolar lattices. 

%\red{Main text}

In this Letter, we develop a general theory of BICs in multipolar lattices and show that off-$\Gamma$ (parametric) BICs, that are usually very sensitive to parameters of the structure, are pinned in the $k$-space to specific directions and they are robust with respect to changes in the resonance frequency of meta-atoms and lattice period as long as the system remains subwavelength. This theory predicts the existence of BICs forming a continuous line in the $k$-space and sets a direct relation between the topological charge of BIC and the asymptotic behaviour of the Q factor in its vicinity.

   \begin{figure}
\begin{center}
\includegraphics[width=1\linewidth]{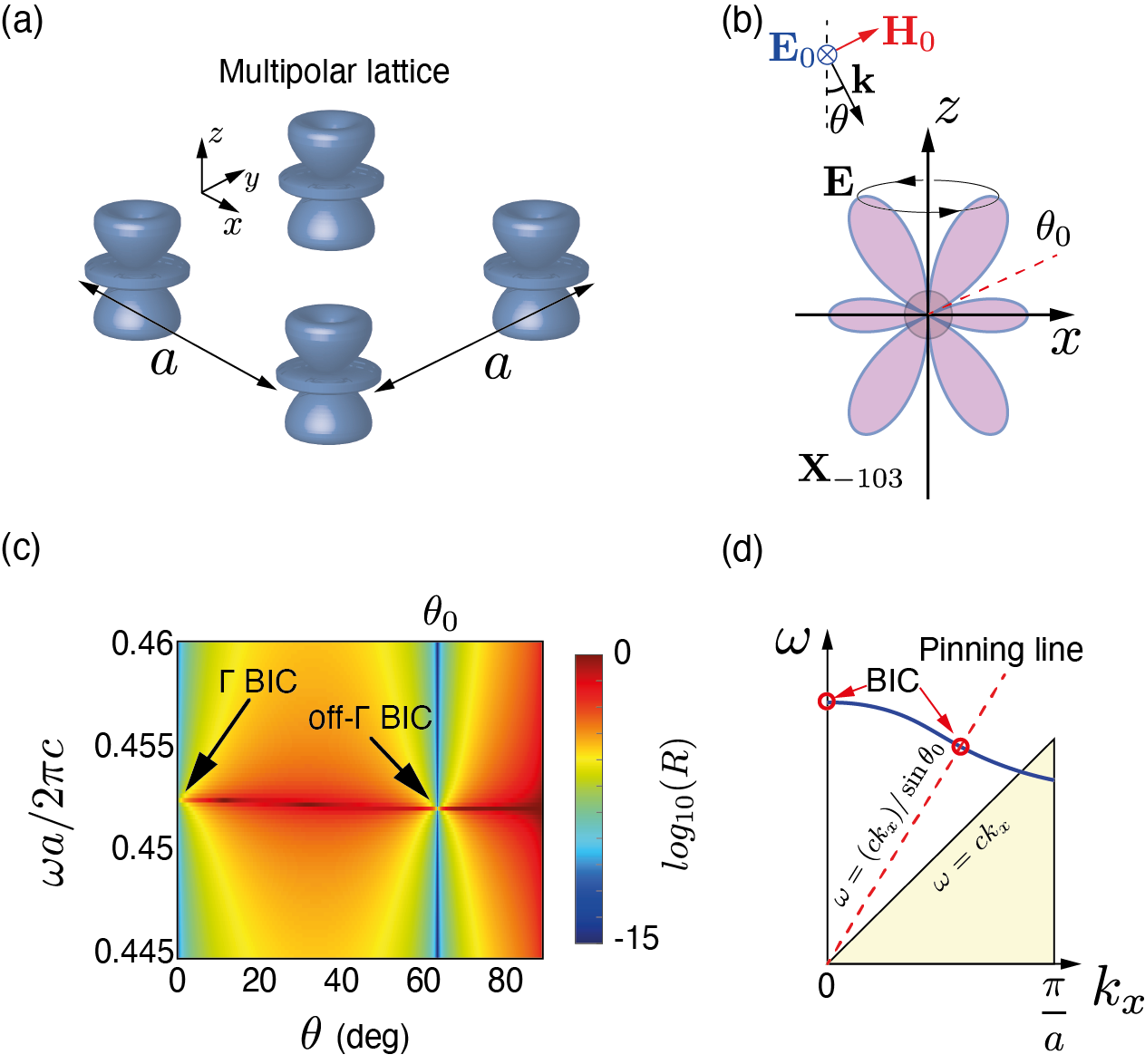}
\caption{(a) Concept of the system. Multipolar lattice with the square unit cell and period $a$ containing a single multipole in the unit cell (electric $\mathbf{Z}_{tm\ell}$ or magnetic $\mathbf{X}_{tm\ell}$).  (b) The far-field radiation pattern plotted for the magnetic octupole scatterer ( $\mathbf{X}_{-103}$) in the $xOz$ plane. The inset shows the polarization of the incident wave. (c) The reflection $R$ from the magnetic octupolar lattice with a square plotted as a function of dimensionless frequency $\omega a/2\pi c$ and angle of incidence $\theta$ for the ration $a/d=1.75$ in logarithmic scale. The magnetic octupole $\mathbf{X}_{-103}$ is described by the T-matrix of a high-refractive-index sphere neglecting all its scattering channels except the magnetic octupolar. The permittivity of the sphere is $\varepsilon=50$ and the diameters is $d$. (d) Schematic bang diagram for the multipolar lattice demonstrating the pinning of the off-$\Gamma$ BIC in the $k$-space.}
\label{fig:concept}
\end{center}
\end{figure}

The main idea of the developed theory is based on the fact that the multipoles arranged into a periodic structure radiate as well as a single multipole, but only to the direction of the open diffraction channel~\cite{chen2019singularities,sadrieva2019multipolar,paniagua2016generalized,ohtaka1979energy}. In other words, the polarization of the plane wave radiated by the metasurface is the same as the polarization of the wave radiated by a single multipole along the same direction. The interaction between the multipoles in the lattice affects only the amplitudes of the multipoles but not their radiation pattern. Therefore, the magnitude of the radiated power into the open diffraction channel is proportional to the directivity of the multipole along this direction. Formally, the far-field of the subwavelength multipolar lattice can be written as (see Supplemental Material)
\begin{equation}\label{eq:farfield}
    \mathbf{E}(\mathbf{r})=\frac{S_{b}}{2 \pi k k_{z}} e^{i \mathbf{k} \mathbf{r}} i^{-\ell} \tilde{D}_{s}\mathbf{Y}_{s}\left(\frac{\mathbf{k}}{k}\right).
\end{equation}
Here $k=\omega/c$ is the wavenumber in the surrounding space that is assumed to be air, $\mathbf{k}$ is the total wavevector of the radiated plane wave,  $k_z=\pm\sqrt{k^2-|\mathbf{k}_b|^2}$, where $\mathbf{k}_{b}$ is the Bloch vector. $S_b$ is the 2D volume of the first Brillouin zone. $\mathbf{Y}_{s}=\{\mathbf{X}_{s}, \mathbf{Z}_{s}\}$ are the vector spherical functions which describe the far-field of each multipole and $\tilde{D}_{s}$ are the coefficients of the multipolar decomposition.  Index $s$ is a set of indices ${p, m, \ell}$, where $l = 0, 1, 2,...$ is the total angular momentum quantum number, and $m = 0, 1,..., \ell$ is the absolute value of the projection of the angular momentum (magnetic quantum number).  Index $p = \pm 1$ defines the parity of $\mathbf{Y}_{s}$ with respect to reflection from the $xz$ plane $(\varphi \rightarrow -\varphi)$ (see Supplemental Material).

% The signs $\pm$ correspond to the upward and downward propagation.  

%\begin{equation}\label{eq:farfield}
%\begin{aligned}
%\mathbf{E}(\mathbf{r})=& E_{0} \sum_{\mathbf{K}, s} \tilde{D}_{s} \frac{V_{b} i^{-\ell}}{2 \pi k} \iint\limits_{-\infty}^{\ \ \ \infty} d \mathbf{k}_{\|} \delta\left(\mathbf{k}_\|-\mathbf{k}_{b}-\mathbf{K}\right) \\
%& \times \frac{e^{i \mathbf{k %r}}}{|\mathbf{k}_{\bot}|}\mathbf{Y}_{s}\left(\frac{\mathbf{k}}{|\mathbf{k}|}\right) .
%\end{aligned}
%\end{equation}
% Here, $\mathbf{k}=\mathbf{k}_\|\pm\mathbf{k}_\bot$ is the wavevector of the scattered wave plane wave, 

The BICs in a multipolar lattice are formed when the directions of all open diffraction channels coincides with the nodal lines of multipoles forming the lattice~\cite{sadrieva2019multipolar,chen2019singularities}. For subwavelength lattices, there is only one open diffraction channel, therefore, each nodal line corresponds to a BIC. The directions of the nodal lines and surfaces are completely defined by the multipole $\mathbf{Y}_s({\mathbf{k}}/{k})$ and they do not depend on the lattice parameters as long as  the multipolar composition of the unit cell is assumed to be conserved.

The developed theory is quite general and can be applied to arbitrary multipolar lattices but to have an illustrative example we consider first a  multipolar lattice with the square unit cell of a period $a$ consisting of single magnetic octupoles ($\mathbf{X}_{-103}$) [see Figs.~\ref{fig:concept}(a) and \ref{fig:concept}(b)]. In numerical simulations, we describe the single multipole by the T-matrix of the high-refractive index sphere with permittivity $\varepsilon$ diameter $d$ neglecting all its scattering channels except one corresponding the multipole under interest. The numerical code is based on the MULTEM package~\cite{newmultem,multem} (see Supplemental Material).        

Figure~\ref{fig:concept}(c) show the reflection $R$ from the octupolar lattice as function of frequency~$\omega$ and angle of incidence~$\theta$. The incident wave is assumed to be $s$-polarized [see Fig.~\ref{fig:concept}(b)]. One can see that the octupolar resonance forms a narrow near flat band featuring weak interaction between the multipoles. The resonance becomes infinitely narrow at angles $\theta_0=0\degree$ and 63.4\degree corresponding to the nodal lines of the octupole manifesting the appearance of the at-$\Gamma$ and off-$\Gamma$ BICs. At these angles, the incident wave does not interact with metasurface and it becomes completely transparent. Therefore, to find the angular position of BICs in the case of arbitrary multipolar lattice, one need to solve a system of the equations $E_\theta=E_\varphi=0$, where $\mathbf{E}$ is the far-field of the multipole defined by Eq.~\eqref{eq:farfield}. For our particular case of magnetic octupoles ($\ell=3$, $m=0$), this system is reduced to one equation $P'_3(\cos\theta)=0$, where $P_3$ is the Legandre polynomial. The roots of this equation exactly gives $\theta_0=0\degree$ and 63.4\degree predicting the angular position of the BICs.             

%We can calculate the angle $\theta_0$ of inclination of the nodal line because we know that 
%\begin{equation}
%    \mathbf{M}_{-103} \sim \frac{d P_3(\cos \theta)}{d \theta}
%\end{equation}
%and we need to solve the equation 
%\begin{equation}
%    \frac{d P_3(\cos \theta)}{d \theta} = 0
%\end{equation}
%with condition $0\leq\theta \leq \pi/2$  and, as a result, $\theta_0 \approx 63.4$ degrees.   

%Fig. 1c demonstrates the reflection map $r(\omega a/2\pi c, \theta)$, where $\Gamma$ and off-$\Gamma$ BICs are formed at the points of intersection of the modes and the lines $\theta = 0$  and  $\theta = \theta_0$, respectively. %The dielectric permitivity of the sphere $\varepsilon_s$ is 50.

Figure~\ref{fig:concept}(d) shows schematically how the BICs in multipolar lattice are formed and pinned to the specific directions of the $k$-space. The blue curve depicts a dispersion band of the metasurface forming by the multipolar resonances. The red dashed curve  $\omega = c k_x/\sin \theta_0$ corresponds to the nodal line of the multipole. The BIC is forming exactly at the crossing of the dispersion surface with the nodal lines of the multipole. The dispersion depends on the polarizability of meta-atoms and lattice parameters while the nodal lines does not depend on them at all that makes off-$\Gamma$ BICs pinned to the specific directions in $k$-space and robust to the variation of lattice parameters.  One should mention that the angular robustness of BIC is ensured by conserving the multipolar composition of the unit cell. The variation of the multipolar compositions results in the migration of BIC within the Brillouin zone.    

%When considering the band diagram $ \omega(k_x)$, we see that the Off-$\Gamma$ BIC is formed as a result of the intersection of the mode and pinning line $ \omega = \dfrac{c k_x}{\sin \theta_0} $ in the systems with an open diffraction channel. This allows us to predict that off-$\Gamma$ BIC may be pinned in the $k$-space (see Fig1 d).

\begin{figure}[t]
\begin{center}
\includegraphics[width=0.85\linewidth]{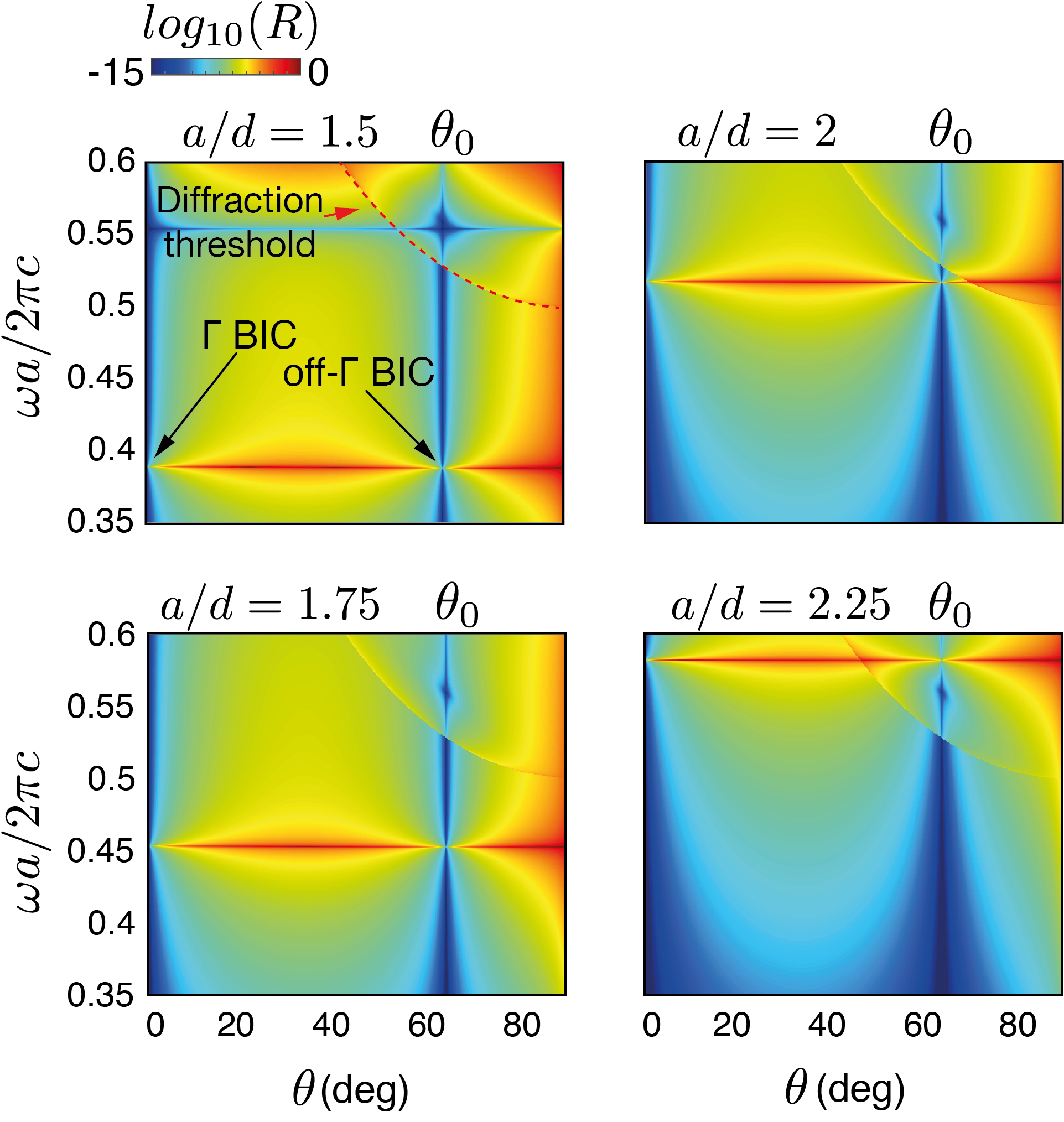}
\caption{The reflection $R$ from the magnetic octupolar ($\ell=3$, $m=0$) lattice with a square unit cell [see Fig.~\ref{fig:concept}(a)] plotted as a function of dimensionless frequency $\omega a/2\pi c$ and angle of incidence $\theta$ for different ratios $a/d$ in logarithmic scale. The period of the structure is $a$.  The magnetic octupole $\mathbf{X}_{-103}$ is described by the T-matrix of a high-refractive-index sphere neglecting all its scattering channels except the magnetic octupolar. The permittivity of the sphere is $\varepsilon=50$ and diameters is $d$. The incident wave is s-polarized [see Fig.~\ref{fig:concept}(a)].}
\label{fig:result1-maps}
\end{center}
\end{figure}

Figure~\ref{fig:result1-maps} shows the reflections maps $R$ for multipolar lattices for different $a/d$ ratios. One can see that variation of $a/d$ ratio affects only the frequency of the off-$\Gamma$ BIC but not its angular position. The increase of $a/d$ from 1.5 to 2.25 results in the blueshift of the band due to the enhancement of the interaction between the multipoles. This can lead to destruction of the BIC if its frequency becomes higher the diffraction threshold, namely, $\omega a /2\pi c>1/(1+\sin\theta_0)$ (see Figure~\ref{fig:result1-maps}, panel for $a/d=2.25$). It is worth mentioning that even for the frequencies higher than the diffraction threshold the radiative losses along the direction of the nodal line is still forbidden and the radiation occurs through other open diffraction channels. One can say that the lack of the radiation along the particular directions is protected by the symmetry of the multipole. A similar effect is observed for the symmetry-protected BIC in the dielectric gratings~\cite{sadrieva2017transition}.

%As already stated, this type of BIC is completely robust to changes of the lattice period $a$ and resonant frequency of the multipoles until the system exceeds the diffraction threshold. In order to see an angular pinning of off-$\Gamma$ BIC it is better to look at the dependence $\omega({\theta})$ for different period of the system $a = 300,350,400 $ nm (see Fig 2). 

\begin{figure}[t]
\begin{center}
\includegraphics[width=1\linewidth]{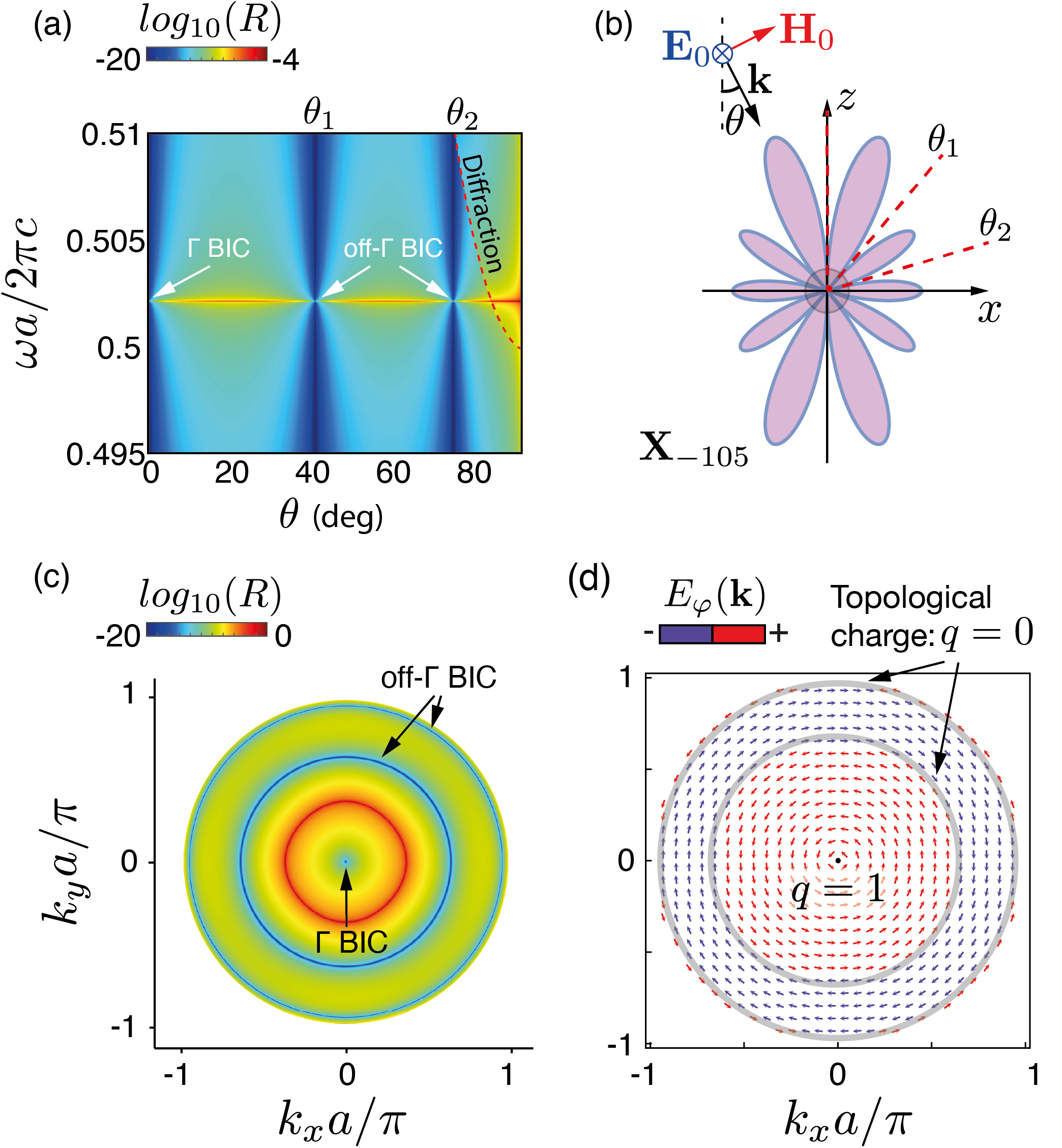}
\caption{(a)  The reflection $R$ from the magnetic multipolar lattice ($\ell=5$, $m=0$) lattice with a square unit cell [see Fig.~\ref{fig:concept}(a)] plotted as a function of dimensionless frequency $\omega a/2\pi c$ and angle of incidence $\theta$ for the ratio $a/d=2$ in logarithmic scale. The period of the structure is $a$.  The magnetic multipole $\mathbf{X}_{-105}$ is described by the T-matrix of a high-refractive-index sphere (permittivity $\varepsilon=220$ and diameter $d$) neglecting  all its scattering channels except the magnetic octupolar. (b) Far-field radiation pattern for $\mathbf{X}_{-105}$.The inset shows the polarization of the incident wave. (c) Reflection map for the same lattice plotted at frequency $\omega a /2\pi c = 0.5$ as a function of the in-plane wavevectors $k_x$ and $k_y$. (d) Polarization map of the magnetic multipolar band ($\mathbf{X}_{-105}$) in the far-field showing the BICs forming a close continues line in the $k$-space.}
\label{fig:result2-two-BICs}
\end{center}
\end{figure}

The higher order multipoles have a larger number of the nodal lines. This results in appearance of several off-$\Gamma$ BICs within a one band. Figure~\ref{fig:result2-two-BICs}(a) shows the reflection map for the metasurface formed by the resonant magnetic multipoles of the fifth order ($\ell=5$; $m=0$). The directivity pattern for such a multipole and its nodal lines corresponding to $\theta_1=40.0\degree$ and $\theta_2=73.3\degree$ are shown in Fig.~\ref{fig:result2-two-BICs}(b). One should mention that as we consider the lattice consisting of the multipoles with the rotational symmetry ($m=0$), the nodal lines are actually nodal surfaces (cones). Thus, the off-$\Gamma$ BICs form a closed continues line~\cite{sadrieva2019multipolar,kostyukov2022ring,bulgakov2019bound,cerjan2019bound}. As the band is nearly flat the BICs form nearly circles in the $k$-space. This can be seen from the reflection map shown in Fig.~\ref{fig:result2-two-BICs}(c).

%Note that there can be more than one off-$\Gamma$ BIC in the first Brullen zone for the higher-order multipolar lattice. For example, the magnetic multipole of fifth order ($\mathbf{M}_{-105}$) has two angles $\theta_1 \approx 40$ and $\theta_2 \approx 73.3$ degrees. Therefore, two off-$\Gamma$ BICs can be formed for this multipolar lattice, which can be clearly seen on the reflection maps $r(\omega a/2\pi c, \theta)$ in Figure 3b.
%However, since  the frequency of the mode should be 
%\begin{equation}
 %   \frac{\omega a}{2\pi c} < \frac{1}{1+\sin{\theta_2}} \approx 0.51
%\end{equation}
%Observing two off-$\Gamma$ BICs is limited by specific materials and geometric parameters in the first Bruellen zone.
%This is necessary to ensure that the off-gamma BIC does not cross the diffraction threshold.
%See the Supplemental Material.
% Dielectric permittivity of particles ($\varepsilon_p$) is 220, dielectric permittivity of medium ($\varepsilon_m$) is 1 (vacuum),  period of the system (a)  is 400 nm, diameter of the sphere (d) is $200$ nm. 

%Since each multipole with $m = 0$ and $l > 2$ has one or several nodal cones in the reciprocal space, we need to pay special attention as this type off-$\Gamma$ BICs forming a continuous line in $k$-space (see Fig 3c and d), rather than a point. Therefore this case of off-$\Gamma$ BICs has non-trivial topological properties.
% It's very interesting from the point of view of the topological nature of this type of off-$\Gamma$ BIC. 

\begin{figure}[t]
\begin{center}
\includegraphics[width=1\linewidth]{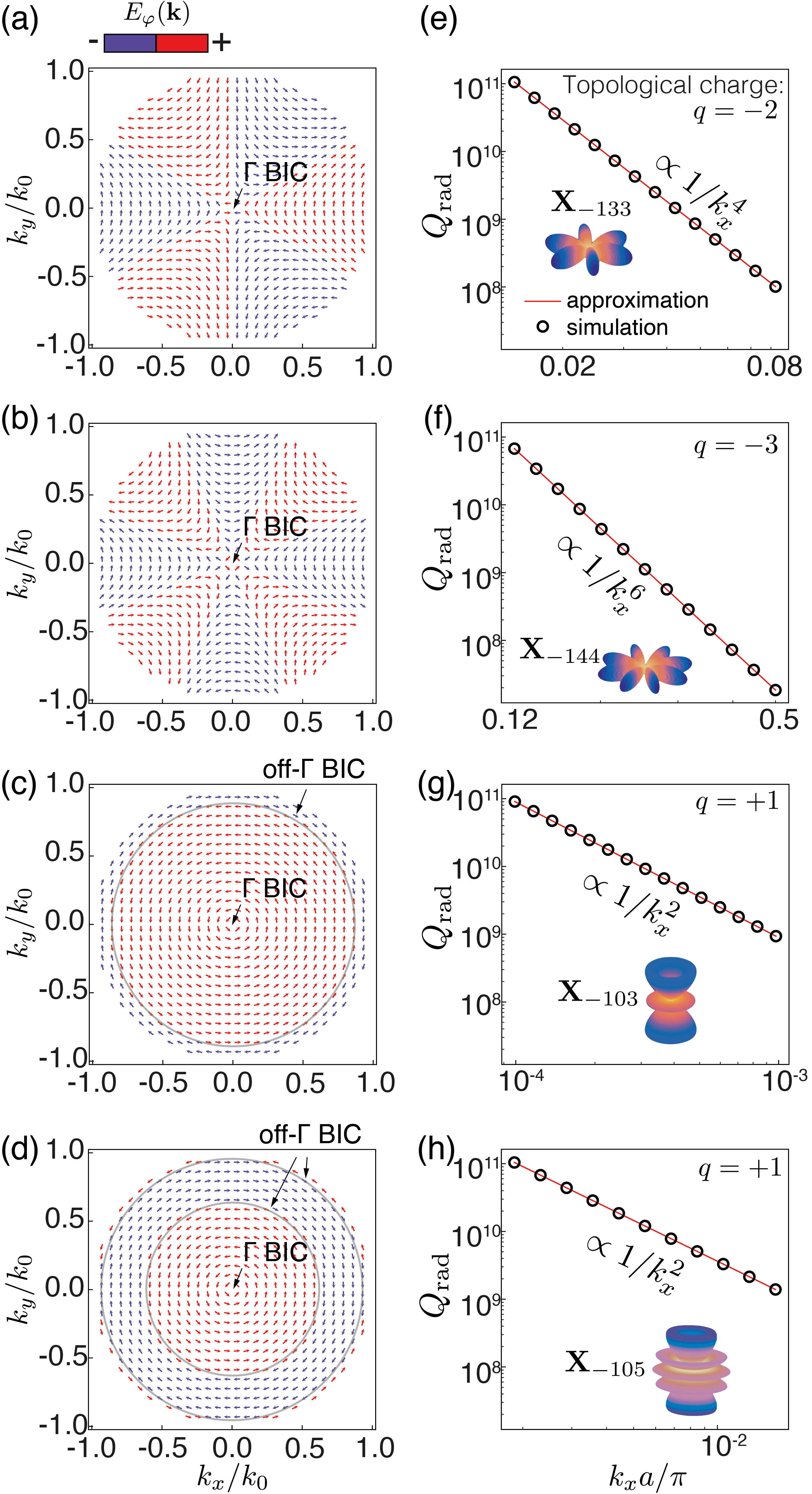}
\caption{(a-d) Far-filed polarization maps of multipolar lattice composed of various multipoles. (e-h) Asymptotic behaviour of $Q_\text{rad}$ in the vicinity of the $\Gamma$-point for plotted for the various multipoles (see insets) found numerically and analytically.}
\label{fig:result3-topology}
\end{center}
\end{figure}

The BICs in periodic photonic structures can be associated with polarization vortices in the far-field which carry a topological charge $q$ that can be calculated as~\cite{zhen2014topological}
\begin{equation}
q=\frac{1}{2 \pi} \oint_{C} d \mathbf{k} \cdot \nabla_{\mathbf{k}} \phi(\mathbf{k}), \quad q \in \mathbb{Z}. 
\label{eq:topologcial_charge}
\end{equation}
Here, $ \phi (\bm{k}) = \arg[E_x(\bm{k}) + iE_y(\bm{k})]$, $E_x$ and $E_y$ the complex amplitudes of the radiated plane waves, and $C$ is a simple counterclockwise oriented path enclosing the singular point.  The reasonable question '{\it Do the BICs forming a continues line in the $k$-space carry some topological charge?}' Figure~\ref{fig:result2-two-BICs}(d) shows the polarization map calculated for the multipolar lattice consisting of the magnetic multipoles ($\ell=5$, $m=0$).      The topological charge $q$ carried by an off-$\Gamma$ BIC forming a closed line in the $k$-space can be defined as the difference between the integrals [Eq.~\eqref{eq:topologcial_charge}] calculated over closed paths inside and outside the BIC-line. The straightforward calculations shows that BICs forming a line in the case space the topological charge $q=0$. Indeed, as one can see from Fig.~\ref{fig:result2-two-BICs}(d), the polarization vortex inside and outside the BIC-line changes only the direction of rotation that does not affect the value of the integral in Eq.~\eqref{eq:topologcial_charge}. This result is in complete accordance with the Poincare-Hopf theorem that reads that only isolated singularities contribute to the total charge~\cite{milnor1997topology}.    

%BICs are vortex centers in the polarization direction of far-field radiation.
% The robustness of these BICs is due to the existence of conserved and quantized topological charges.
%Vortices on the map of the electric field $\mathbf{E}$ in the $k$-space are characterized by their topological charges. Here, the topological charge ($q$) carried by a BIC is defined as
%\begin{equation}
%q=\frac{1}{2 \pi} \oint_{C} d \mathbf{k} \cdot \nabla_{\mathbf{k}} \phi(\mathbf{k}), \quad q \in \mathbb{Z}
%\end{equation}
%which describes how many times the polarization vector winds around the vortex center. Here, $ \phi (\bm{k}) = arg[E_x(\bm{k}) + iE_y(\bm{k})]$  is the angle of the polarization vector, and C is a closed simple path in k space that goes around the BIC in a counterclockwise direction \cite{zhen2014topological}. 

%Thus, we see that the topological charge $q$ of off-$\Gamma$ BIC is 0 for multipoles with $m = 0$ and $l > 2$.  This is shown on the map of the vector field $E(\bm(k))$ and the reflection map which depends on $k_x$ and $k_y$ for fixed dimensionless frequency $\omega a /2\pi c = 0.502$  in the first Brillouin zone (BICs is marked with a dotted line in the figures 3 c and d).

In theory, BIC can carry arbitrary high topological charge~\cite{zhen2014topological}. Such BICs are quite perspective as they can be more robust to imperfections of the structures and demonstrate higher Q factors~\cite{jin2019topologically}. Nevertheless, up to date, all suggested designs of the photonic crystals and metasurfaces support the BICs the maximal topological charge $|q|=2$~\cite{yoda2020generation,dyakov2021photonic}. Quasi-crystals due to a high-order rotational symmetry can support high-Q leaky resonances with polarization singularities of a large topological charge~\cite{che2021polarization}. However, the observation of the genuine BICs with high topological  charges is still a challenge. However, the multipolar lattices can naturally support them. Indeed, one can see from Eq.~\eqref{eq:farfield} that the polarization structure in the vicinity of the $\Gamma$-point entirely succeeds the far-field polarization structure of the multipole including all polarization singularities~\cite{chen2019singularities}. Therefore, to construct the BIC with high topological charge in the multipolar lattice, one needs to take a multipole forming a vortex along the vertical direction with a large topological charge. For example, multipolar lattices consisting of the electric or magnetic multipoles with $\ell=m\geq2$ support the symmetry-protected BICs with topological charge $q=1-\ell$. Figures~\ref{fig:result3-topology}(a) and \ref{fig:result3-topology}(b) show the polarization map for the multipolar lattices comprised by the magnetic multipoles with $\ell=m=3$ and $\ell=m=4$, respectively. Applying Eq.~\eqref{eq:topologcial_charge} to these maps, one can get the topological charges $q=-2$ and $q=-3$. Taking higher order multipoles one can get BICs with arbitrary high negative topological charges.       

Asymptotic behaviour of the Q factor in the vicinity of BIC is quite important from the practical point of view as it allows to predict the robustness of the BIC to the structural imperfections and the dependence of the radiative Q factor on the size of the sample~\cite{sadrieva2019experimental,jin2019topologically}. For the multipolar lattices, the dependence of the radiative Q factor on the Bloch wavevector $\mathbf{k}_b=(k_x,k_y,0)$ can be found in analytically for the considered multipolar lattices using the following definition $Q = \omega {W_\text{st}}/{P_\text{rad}}$,where $W_\text{st}$ is the energy stored in the unit cell and $P_\text{rad}$ is the power radiated by the unit cell. Assuming that the multipole is described  by the T-matrix of a high-refractive-index sphere and all the energy is stored inside the sphere,  one can show that the asymptotics of the Q factor for BIC in the vicinity of the $\Gamma$-point is completely defined by the topological charge of the BIC:  
\begin{equation}
    Q \sim \frac{1}{|\mathbf{Y}_{s}({\bm{k}}/{k})|^2} \sim \frac{1}{|\mathbf{k}_b|^{2|q|}}.
\end{equation}        
The exact analytical expression is quite cumbersome and we provide it in Supplemental Material. Figures~\ref{fig:result3-topology}(e)-\ref{fig:result3-topology}(h) show the asymptotics of the Q factor in the vicinity of the $\Gamma$-point for different multipolar lattices obtained numerically and analytically. The corresponding polarization maps are shown in Figs.~\ref{fig:result3-topology}(a)-\ref{fig:result3-topology}(d).

% \red{Summary}

In conclusion, we analyzed at-$\Gamma$ and off-$\Gamma$ BICs in the subwavelength multipolar lattices -- metasurfaces whose unit cell contains only one resonant multipole. Technically, we describe the single multipole by the T-matrix of a high-refractive-index sphere neglecting all its scattering channels except one corresponding to the multipole under interest. We predict that the off-$\Gamma$ BICs in such lattices are pinned to the specific directions in the $k$-space and completely robust against changes of the resonant frequency of the multipole and lattice parameters as long as the structure remains subwavelength. The analyzed BICs are protected by the symmetry of the multipoles. We show that if the multipole has a rotational symmetry with respect the the direction normal to the metasurface, the BIC forms a continues closed line in the $k$-space. The topological charge for such BICs are zero. We reveal that the multipolar lattice can host the at-$\Gamma$ BICs with a high topological charge $|q|>2$ and set a direct analytical relation between the asymptotics of Q factor in the vicinity of the BICs and their topological charge. One should necessarily mention that the multipolar lattices containing only one multipole in the unit cell are an only approximation that works well only in the vicinity of the resonance when one multipole dominates over others. The more accurate description requires accounting for other multipoles.  We believe that the developed theory and predicted effects are an important step in understanding of the physics of bound states in the continuum and their engineering in periodic photonic structures. 

%Despite the fact that the multipolar lattice do not give a direct recipe for designing the unit cell,

The authors thank K. Koshelev, D.  Maksimov for useful discussions. The analytical results and simulation were obtained with the supported of the Russian Science Foundation project \#20-72-10141. S.\,G., K.\,F., and A.\,B acknowledge the "BASIS" Foundation and the Priority 2030 Federal Academic Leadership Program.

\newpage
\bibliography{reference-1}

\end{document}